\documentclass[a4paper,10pt]{article}

\begin{document}
\title{{\bf {Currents in System of Noisy Mesoscopic Rings}}}
\author{J. Dajka, J. {\L}uczka, M. Szopa, and E. Zipper\\
Institute of Physics, University of Silesia,\\ 40-007 Katowice,
Poland } \maketitle

\begin{abstract}
A semi-phenomenological model is proposed to study dynamics and
steady-states of magnetic fluxes and currents in mesoscopic rings and
cylinders at non-zero temperature. The model is based on a Langevin equation
for a flux subject to zero-mean thermal equilibrium Nyquist noise. Quenched
randomness, which mimics disorder, is included via the fluctuating parameter
method. In the noiseless case, the stability threshold (critical
temperature) exists below which selfsustaining currents can run even if the
external flux is switched off. It is shown that selfsustaining and
persistent currents survive in presence of Nyquist noise and quenched
disorder but the stability threshold can be shifted by noise.
\end{abstract}

PACS number(s): 05.40.-a, 73.23.Ra, 02.50.Ey



\section{Introduction}

Quantum phenomena manifested at the mesoscopic level have attracted much
experimental and theoretical attention. Phase coherence and persistent
currents can be mentioned as examples. Persistent currents were predicted as
early as 1938 \cite{hund}  and have been observed experimentally only since
1990 \cite{levy}. In the paper we analyze the steady states of magnetic
fluxes and currents in a mesoscopic system subject to dissipation and
fluctuations \cite{kog}. The system consists of a set of concentric one
dimensional rings stacked along a certain axis. It is known that  a small
metallic ring threaded by a magnetic flux displays a persistent current,
signifying quantum coherence of electrons (called coherent electrons)  and
it is a direct manifestation of phase-coherent electron motion and the
Aharonov-Bohm effect at the mesoscopic level. Moreover, it has been shown
\cite{koh} that in such a system selfsustaining currents run even if the
external flux is switched off. At temperature $T=0$, the system is in the
ground state and only coherent electrons exist. Then the persistent current
flows without dissipation.  At temperature $T>0$ the amplitude of the
persistent current run  by coherent electrons decreases and some electrons
become "normal"  (i.e. non-coherent).  The motion of normal electrons  is
random and their flow is dissipative. Under some conditions, coherent
conduction and normal conduction coexist, resulting in dissipation of a
total current. It was confirmed experimentally in \cite{le} that mesoscopic
rings connected to a current source presented an Ohmic resistance which was
not zero.

We introduce a semi-phenomenological model describing the flux and current
in such a system, which takes into account both dissipation and
fluctuations. The approach is based on the notion of the two fluid model
\cite{twofluid} of normal and coherent electrons formulated as a Langevin
equation with a noise term. In the system at temperature $T>0$ there are
various sources of noise and fluctuations. There are so-called universal
conductance fluctuations \cite{han} that arise from the random quantum
interference between many electron paths which contribute to the conductance
in the diffusive regime. These fluctuations decay algebraically with
temperature and can be neglected at higher temperatures \cite{han}.
Inelastic transitions in the ring cause another kind of fluctuations.
However, they do not destroy persistent currents but decrease their
amplitude \cite{bu85}. There is also a part of the current noise which is
called shot noise \cite{kog}, the spectral density of which is proportional
to mean current. This noise can be reduced by increasing the size of rings
\cite{land}. Thermal motion of charge carriers in any conductor is a source
of random fluctuations of current which is called Nyquist noise \cite{kog}.
This thermal equilibrium noise is universal and exists in any conductor,
irrespective of the type of conduction and is analogous to noise driving the
motion of Brownian particles. Moreover, this noise increases with
temperature. Therefore at relatively high temperatures and relatively large
rings, universal conductance fluctuations and shot noise can be neglected,
and only Nyquist noise can play an important role. This is the case
considered in the paper. We assume that coherent and normal electrons
coexist \cite{twofluid} and that normal electrons are subjected to  Nyquist
noise. This noise generates the flux fluctuations which indirectly influence
persistent currents run by coherent electrons. Our main goal is to answer
the question whether persistent and selfsustaining currents survive in the
presence of dissipation and fluctuations. In classical physics dissipation
and fluctuations may often be described by introducing random forces into
evolution equations. In quantum physics, the inclusion of fluctuations
requires more care because quantum systems are described by Hamilton
operators and it ensures conservation of energy. In the paper we combine a
classical Langevin equation with noise term and with terms of a quantum
origin. Our model is minimal in the sense that in two limiting cases it
reduces to the well-established models of the quantum persistent current of
coherent electrons and the classical Nyquist current of normal electrons.
The approach used could be justified in a more elegant way applying the
methods of the thermo-field dynamics \cite{ume}.

The paper is structured as follows. In Sec. II, we describe the model based
on the concept of two fluids and introduce the Langevin equation for the
magnetic flux, which include Nyquist fluctuations. Its detailed analysis is
carried out in Sec. III. In this section, influence of quenched
disorder on the system is analyzed as well.  Conclusions are made in Sec. IV.
Finally, in the Appendix, we present the Nyquist relation from which the
intensity of thermal fluctuations is determined.

\section{ Two fluid model of electronic transport}

Persistent currents in conducting structures with topology of the ring are
an example of a quantum-interference phenomenon in which the phase coherence
of the electron wave function affects equilibrium and transport properties
of mesoscopic systems. At zero temperature, $T=0$, the current in a quasi
one-dimensional mesoscopic ring threaded by the magnetic field of the
Bohm-Aharonov type is paramagnetic \cite{cheng}
\begin{equation}  \label{koha1}
I_{even}(\phi) = I_0 \sum_{n=1}^{\infty} A_n \sin\left(\frac{2n\pi\phi}{%
\phi_{0}}\right)
\end{equation}
in the case of an even number of coherent electrons and diamagnetic \cite%
{cheng} 
\begin{equation}  \label{koha2}
I_{odd}(\phi) = I_{even}(\phi + \phi_0 /2)
\end{equation}
%
in the case of an odd number of coherent electrons. Here, $\phi$ is the
magnetic flux, $\phi_0:=h/e$ is the flux quantum, the amplitudes
\begin{equation}  \label{amp1}
A_n = \frac{2}{\pi n} \cos(nk_{F}l_x),
\end{equation}
the unit current $I_0=h eN_{e}/(2 l_{x}^{2} m)$, $k_F$ is the Fermi
momentum,  $l_x$ is the circumference of the ring and $N_e$ is the number of
coherent electrons in the ring.

Now, let us consider a collection of rings, individual current channels,
which form a cylinder. There are $N_z$ channels in direction of the cylinder
axis and $N_r$ in the direction of the cylinder radius. We assume that the
thickness of the cylinder wall is small if compared with the radius. Because
of the mutual inductance between rings, the current in one ring induces flux
in other rings. In turn, the flux induces a current, and so on. We will
analyze the effect of the mutual inductance among the rings. We assume that
the rings are not contacted. So, there is no tunneling of electrons among
the channels and the charge carriers moving in the different rings are
independent. It has been shown in \cite{phrevB} that the effective
interaction between the ring currents, when taken in the selfconsistent mean
field approximation, results in the magnetic flux $\phi =LI_{tot}$ felt by
all electrons, where $L$ is the cylinder inductance and $I_{tot}$ is the
total current in a cylinder. For a cylinder of the radius $r$ and the height
$l_{z} $ the inductance is \cite{bloch}
\begin{equation}
L=\mu _0\frac{\pi r^2}{l_z} ,  \label{L}
\end{equation}
where $\mu_0$ is the permeability of the free space. At temperature $T>0$,
the current $I_{coh}(\phi ,T)$ of the coherent electrons in a set of $%
N=N_r\times N_z$ current channels forming the cylinder is either \cite{koh}
\begin{equation}  \label{Iev}
I_{coh}(\phi ,T)=I_{even}(\phi ,T)  \nonumber \\
=NI_0\sum_{n=1}^\infty A_n(T)\sin \left( \frac{2n\pi \phi }{\phi _0}\right)
\end{equation}
for an even number of coherent electrons in each single channel or
\begin{equation}  \label{Iod}
I_{coh}(\phi ,T)=I_{odd}(\phi ,T)=I_{even}(\phi +\phi _0/2,T)
\end{equation}
for an odd number of coherent electrons. The amplitude
\begin{eqnarray}  \label{Am}
A_n(T)=\frac{4T}{\pi T^{*}}\frac{\exp (-nT/T^{*})}{1-\exp (-2nT/T^{*})}\cos
(nk_Fl_x).
\end{eqnarray}
The characteristic temperature $T^{*}$ is given by the relation $%
k_BT^{*}=\Delta _F/2\pi ^2$, where $k_B$ is the Boltzmann constant and $%
\Delta _F$ is the energy gap at the Fermi surface. For temperatures $T < T^*$
the coherent current flows in such a cylinder without dissipation but its
amplitude (\ref{Am}) is reduced \cite{cheng}. On the other hand, at
temperature $T>0$, normal electrons occur and their flow is dissipative. The
motion of normal electrons is random, like the motion of electrons in a
normal conductor and it generates random currents.

The current coming from the normal electrons can be induced by e.g. the
change of the magnetic flux $\phi $. According to the Lenz's rule and the
Ohm's law one gets \cite{jackson}
\begin{equation}
RI_{nor}(\phi)=-\frac{d\phi }{dt},  \label{nor}
\end{equation}
where $R$ is the effective resistance of the system \cite{bu85}.

At temperature $T>0$, coherent and normal electrons can coexist resulting in
dissipation of the total current $I_{tot}$ which, in the absence of
fluctuations, is a sum of the persistent and normal currents,
\begin{equation}
I_{tot}(\phi ,T)=I_{coh}(\phi ,T)+I_{nor}(\phi) .  \label{tot}
\end{equation}

The magnetic flux is related to the total current via the expression
\begin{equation}
\phi = \phi _{ext} + LI_{tot}(\phi ,T),  \label{LI}
\end{equation}
i.e. it is a sum of the external flux $\phi _{ext}$ and the flux coming from
the currents.

Combining (\ref{tot}), (\ref{nor}) and (\ref{LI}) yields the equation
\begin{equation}  \label{det}
\frac{1}{R}\frac{d\phi}{dt} = -\frac{1}{L}(\phi-\phi_{ext}) + I_{coh}(\phi,
T).
\end{equation}
It is known that the current-flux characteristics for the coherent electrons
is extraordinary sensitive to a change of parity of the coherent carriers
number \cite{cheng}. In order to take into account the possible difference
of parity in the rings we consider the current of coherent electrons as the
average $I^{av}_{coh}(\phi,T)=pI_{even}(\phi,T)+(1-p)I_{odd}(\phi,T)$, where
$p \in [0, 1]$ is the probability of the even number of coherent electrons
in a given channel.

The term describing current fluctuations should also be included in the
evolution equation (\ref{det}). Within our model, the only source of
fluctuations is the equilibrium Nyquist normal current noise generated by
the resistance $R$. The correlation function of this source of fluctuations
is assumed to be given by the Nyquist relation. It follows from the Appendix
that Eq. (\ref{det}) should be modified to the following form
\begin{eqnarray}  \label{lang1}
\frac{1}{R}\frac{d\phi}{dt} = -\frac{1}{L}(\phi-\phi_{ext}) +
I_{coh}^{av}(\phi, T) + \sqrt{\frac{2 k_BT}{R}}\; \Gamma (t),
\end{eqnarray}
where $\Gamma (t)$ is Gaussian white noise (see the Appendix). This equation
takes the form of a classical Langevin equation and is our basic evolution
equation.


Let us introduce dimensionless variables.  In the Langevin equation (\ref%
{lang1}), the basic quantity is the magnetic flux $\phi = \phi (t)$. The
natural unit of the flux is the flux quantum $\phi _0 = h/e$. Accordingly,
the flux is scaled as $x= \phi /\phi _0$. To identify the characteristic
time $\tau_0$, let us consider a particular case of (\ref{lang1}), namely,
when the persistent current and the external flux are zero. Then
\begin{eqnarray}  \label{dv}
\frac{d\phi}{dt} = -\frac{R}{L}\phi + \sqrt{2R k_BT}\; \Gamma (t).
\end{eqnarray}
From this equation it follows that the mean value
\begin{eqnarray}  \label{vt}
\langle \phi(t) \rangle = \langle\phi(0)\rangle\,\exp(-t/\tau_0),
\end{eqnarray}
where
\begin{eqnarray}  \label{taul}
\tau_0 = L/R
\end{eqnarray}
is the relaxation time of the averaged normal current. Therefore time is
scaled as ${\tilde t}=t /\tau_0$. In this case, Eq. (\ref{lang1}) can be
transformed into its dimensionless form
\begin{eqnarray}  \label{aa}
\dot x = -V^{\prime}(x) + \sqrt{2D} \; \widetilde{\Gamma}({\tilde t}),
\end{eqnarray}
where the dot denotes a derivative with respect to the rescaled time ${%
\tilde t}$ and the prime denotes a derivative with respect to $x$. The
generalized potential
\begin{equation}  \label{pot}
V(x) = V(x, \lambda, i_0, p, T)=\frac{1}{2}x^2-\lambda x-i_0 F(x, p, T),
\end{equation}
where $\lambda = \phi_{ext}/\phi_0 $ is the rescaled external flux. The
prefactor $i_0 = N L I_0/ \phi_0$ is a coupling constant characterizing the
interaction between ring currents (it is the rescaled amplitude of the flux
created by the current - it leads to selfsustaining currents).  The function
\begin{equation}  \label{F}
F(x)= F(x, p, T) = \int f(x, p, T)dx
\end{equation}
characterizes the coherent electrons and
\begin{equation}  \label{fx1}
f(x, p, T) = pf_{even}(x, T)+(1-p)f_{odd}(x,T),
\end{equation}
where
\begin{equation}
f_{even}(x,T)= \sum_{n=1}^{\infty} A_n(T) \sin(2n\pi x)
\end{equation}
and 
\begin{equation}  \label{koh2}
f_{odd}(x, T) = f_{even}(x + 1/2, T).
\end{equation}
The dimensionless intensity $D$ of rescaled Gaussian white noise $\widetilde{%
\Gamma}({\tilde t})\equiv \sqrt{\tau_0}\;\Gamma(\tau_0 {\tilde t})$ is a
ratio of thermal energy to the elementary energy stored up in the
inductance,
\begin{eqnarray}  \label{DV}
D = \frac{1}{2}k_BT/\epsilon _0, \qquad \epsilon _0 :=\frac{\phi _0^2}{2L}.
\end{eqnarray}
The rescaled Gaussian white noise $\widetilde{\Gamma}({\tilde t})$ has the
same moments as in (\ref{gauss}). Let us notice that the resistance $R$ does
not occur explicitly in the rescaled equation (\ref{aa}).

Let us evaluate the order of magnitude of the parameters appearing in our
equations. The rescaled coupling constant
\begin{equation}  \label{numi}
i_{0}=\frac{\mu _{0}e^{2}N}{8\pi m_{e}}\frac{N_{e}}{l_{z}}.
\end{equation}
We assume that the cylinder has the radius $r=3\cdot 10^{4}\mathring{A}$ and
the height $l_{z}=100\mathring{A}$. It consists of a set of $N\sim 50$
current channels \cite{Riedel} in a wall of width much smaller than the
radius. If the number of electrons in each channel is $N_{e}\sim 2\cdot
10^{5}$ then $i_{0}\sim 1$. The energy gap at the Fermi surface $\Delta
_{F}=\hbar ^{2}N_{e}/(2m_{e}r^{2})$ gives the rescaled noise amplitude
\begin{equation}  \label{numd}
\frac{k_{B}T^{\ast }}{2\epsilon _{0}}=\frac{\mu _{0}e^{2}}{16\pi ^{3}m_{e}}%
\frac{N_{e}}{l_{z}}.
\end{equation}
For the above values of parameters the diffusion coefficient $D \sim
0.001T/T^{\ast }$. Below, unless stated otherwise, the parameter are fixed
so that $i_{0}= 1$ and $D=0.001T/T^{\ast }$.

\section{Analysis}

In this section the properties of system described by Eq. (\ref{aa}) are
analyzed. We consider in details three special cases. In the deterministic -
noiseless - case, we neglect the influence of  Nyquist noise. It is a
justified approximation for very small intensity of noise. In the second
case, we include  Nyquist noise, i.e. we analyze the full version of (\ref%
{aa}). In the third case, we extend the model assuming that the coupling
parameter $i_0$ is a random variable. It can mimic uncorrelated, quenched
disorder of the rings.

As follows from (\ref{LI}), the current $I_{tot}$ is linearly related to the
magnetic flux $\phi$ (or the rescaled flux $x$). In a consequence, the
properties and behavior of the current are identical to the properties and
behavior of the magnetic flux. Therefore, below we discuss mostly the
dependence of the magnetic flux on the parameters of the model.

\subsection{Noiseless case}

First, let us consider the deterministic case of the Langevin
stochastic  equation (\ref{aa}) formally neglecting the Nyquist
noise term $\tilde{%
\Gamma}({\tilde t})$, i.e.,
\begin{equation}  \label{evol}
\dot x= -V^{\prime}(x).
\end{equation}
The stationary solutions $x_s$ of (\ref{evol}), for which $\dot x_s=0$,
correspond to extrema of the generalized potential (\ref{pot}),
\begin{equation}  \label{stat}
V^{\prime}(x_s) = x_s - \lambda - i_0 f(x_s, p, T)=0.
\end{equation}
The solutions $x_s$ of the gradient differential equation (\ref{evol}) are
stable provided they correspond to a minimum of the generalized potential (%
\ref{pot}) and they are unstable in the case of a maximum \cite{hale}. In
the following we investigate properties of solutions $x_s$ with respect to
four independent parameters: the temperature $T$, the coupling constant $i_0$
which characterizes the mean-field interaction between rings, the
probability $p$ of the occurrence of the channel with an even number of
coherent electrons and the external flux $\lambda$.

\subsubsection{$T$ and $i_0$-dependence}

The dependence of the potential (\ref{pot}) on the temperature for $\lambda
=0$, $i_{0}=1$ and the probability $p=1/2$ is shown in Fig. 1.

 \begin{figure}
\vspace{6cm} \includegraphics{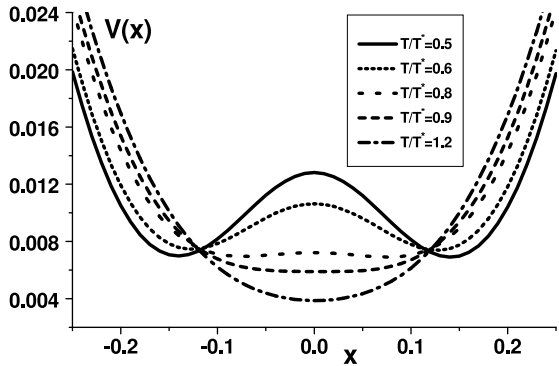} \caption{Dimensionless
generalized potential $V(x)$ given by Eq. (17) is shown as a
function of the dimensionless magnetic flux $x$ for several values
of the scaled temperature $T/T^*$. The scaled amplitude $i_0=1$
and scaled external magnetic flux $\lambda =0$.}
\end{figure}

 In high
temperatures, only one stable solution, corresponding to zero stationary
flux $x_{s}=0$ and zero current, exists. If temperature decreases, a
bifurcation occurs - the potential becomes bistable and two non-zero
symmetric minima appear at $x_{s}=\pm x_{m}$. They correspond to two stable
stationary solutions. Physically, it means that below some critical
temperature $T_{c}$ the {\em spontaneous flux} \cite{self} appears and
non-zero stationary current flows in the system. This critical temperature $%
T_{c}$ is defined by the condition that $V^{\prime \prime }(x_{s}=0)=0$. The
corresponding diagram is shown in Fig. 2.

\begin{figure}
\vspace{6cm} \includegraphics{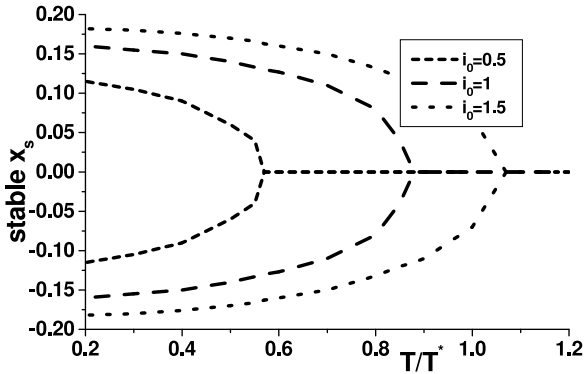} \caption{Bifurcation of the
stable stationary magnetic flux $x_s$ determined by Eq. (26) is
depicted vs temperature $T/T^*$ for several values of $i_0$ and
fixed external flux $\lambda =0$. }
\end{figure}

The phenomenon is analogous to the {\em continuous phase
transition} in macroscopic systems, and appears here
as a result of the interaction of ring currents. The central maximum $%
x_{s}=x_{M}=0$, corresponds to the unstable stationary solution of (\ref%
{evol}).

More generally, one can notice that the stationary solutions occur where the
linear part $x-\lambda $ of (\ref{stat}) is equal to its periodic part $%
i_{0}f(x,p,T)$. In the limit of $i_{0}\rightarrow 0$ (very small, or no
interaction of ring currents), regardless $T$, the only stationary solution
of (\ref{stat}) is the external flux $x_{s}=\lambda $. For intermediate $%
i_{0}$ (typical interaction of mesoscopic rings) two stable non-zero
stationary states can exist below $T_{c}$ and this number of solutions is
preserved in the limit $T\rightarrow 0$. As one can infer from (\ref{pot})-(%
\ref{koh2}), decreasing temperature enhances the periodic part of (\ref{stat}%
) but only to a maximal value defined by $T\rightarrow 0$. Further
enhancement of the periodic part is possible only by increasing the coupling
constant $i_{0}$. As a result of that, the critical temperature $T_{c}$
increases with $i_{0}$. Therefore, if $i_{0}$ is sufficiently large (very
strong interaction of rings), even more stationary states can occur. The
number of stationary states below $T_{c}$ and for $p=1/2$ can, in general,
be equal to $4k-1,$ $(k=1,2...)$ but only $2k$ of them of stable states.
Lowering the temperature below $T_{c}$ results then in a cascade of
bifurcations. The first bifurcation takes place at $T=T_{c}$. With further
lowering the temperature at $T=T_{c_{1}}<T_{c}$ two additional pairs of
stationary solutions appear and so on. There is one metastable and one
unstable solution in every pair. The metastable solutions correspond to the
so called {\em flux trapped} in the cylinder. Notice that in the limit $%
T\rightarrow 0$ and typical $i_{0}>0$ there are always spontaneous flux
solutions whereas the flux trapped solutions can be obtained only for
sufficiently large $i_{0}$.


\subsubsection{The $p$-dependence}

In the following part of this section the temperature is set below $T_c$. If
the probability $p=1$ we have an even number of coherent electrons and
paramagnetic current in each channel. The potential possesses two minima
corresponding to spontaneous fluxes. Decreasing $p$, the probability $1-p$
of finding odd channels with diamagnetic currents increases and the
spontaneous flux solutions $x_s$ decrease to coalesce finally into a single
absolutely stable solution at $x_{s}=0$. The ratio $p$ at which the
coalescence occurs decreases with decreasing temperature. Now, for
sufficiently large $i_0$, five stationary states exist. Note that apart from
the stable fluxless solution $x_s=0$ there are two metastable solutions at $%
\frac{1}{2}< \vert x_s \vert <1$ and two unstable solutions. The metastable
solutions correspond to the flux trapped in the cylinder. In realistic
devices they are hardly accessible due to the value of the necessary
parameters. Further the discussion is limited to the case when $p=1/2$.


\subsubsection{The $\protect\lambda$-dependence}

The dependence of the stationary solutions $x_{s}$ on the external flux $%
\lambda $ is shown in Fig. 3.

\begin{figure}
\vspace{6cm} \includegraphics{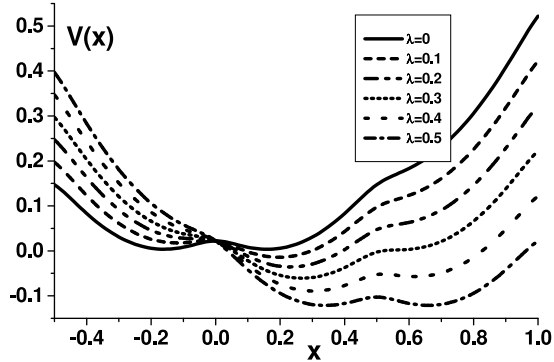} \caption{ Generalized potential
(17) for several values of the  external flux $\lambda$, fixed
$i_0=1$ and  $T/T^* =0.2$. }
\end{figure}

There are three different types of the generalized potential.
First is a symmetric double well potential which appears for
$\lambda =k/2$ with integral $k$. The stable solutions $x_{s}$ are
then always around the external flux $0<|x_{s}-\lambda
|<\frac{1}{2}$. For the values of $\lambda $ close but not equal
to $k/2$ the solutions remain in that range but the double-well
potential becomes asymmetric - one of the stable solutions becomes
metastable. For the values of external flux far from half integer
values $k/2$ one obtains the potential with only a single stable
solution. All the mentioned types of potentials are accessible for
$0\leq \lambda <1/2$ indicating a kind of the 'structural
periodicity' with respect to the external flux.

An interesting feature of the $x-\lambda $ characteristic is the occurrence
of the hysteresis loop (Fig.4).

\begin{figure}
\vspace{6cm} \includegraphics{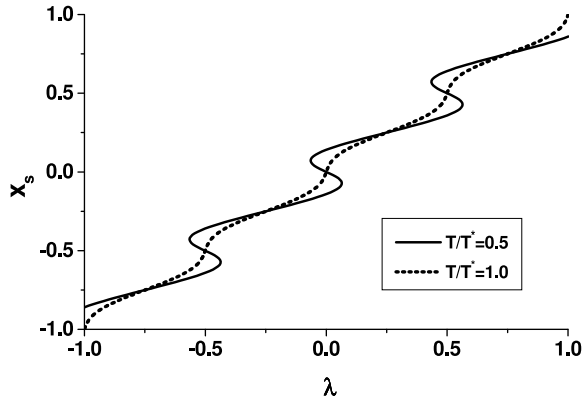} \caption{ The stationary
magnetic flux $x_s$ determined by  Eq.(26) vs  external flux
$\lambda$ and two distinct temperatures. For $T/T^*=0.5$ the part
of the graph with the
negative slope corresponds to unstable $x_s$. The dependence of $x_s$ on $%
\lambda$ is in this case hysteretic. The amplitude $i_0=1$.}
\end{figure}

 With increasing $\lambda , $at its certain
value, the system undergoes discontinuous jump of $x$. Decreasing then the
value of $\lambda $, the opposite jump of $x$ occurs at lower $\lambda $
producing a hysteresis loop. It is a hallmark of the {\it first order phase
transition}. The transition can occur only below the critical temperature $%
T_{c}$. Due to the 'structural periodicity' the hysteresis loop is
repeated with the period $\lambda =1/2$ what results in the formation of a
family of loops.


\subsection{Model with Nyquist noise}

Noise and fluctuations are ubiquitous in real systems and idealization of
the noiseless systems is sometimes not justified. In the following, we will
focus on the system (\ref{aa}) subjected to Nyquist noise. From the
mathematical point of view, the Langevin equation (\ref{aa}) defines a
Markov diffusion process. Its probability density $p(x, {\tilde t})$ obeys
the Fokker-Planck equation in the form
\begin{equation}  \label{fp0}
\frac{\partial }{\partial {\tilde t}}p(x,{\tilde t})= \frac{\partial}{%
\partial x}V^{\prime}(x)p(x, {\tilde t}) + D\,\frac{\partial^2}{\partial x^2}%
p(x, {\tilde t})
\end{equation}
with the natural boundary condition $\lim_{|x|\rightarrow \infty}p(x,{\tilde %
t})=0$. The stationary solution $p_s(x)$ is asymptotically stable \cite%
{lasota} and takes the form
\begin{equation}  \label{s0}
p_s(x)=N_0 \mbox{e}^{-V(x)/D}
\end{equation}
with a normalization constant
\begin{equation}  \label{N0}
N_0^{-1} = \int_{-\infty}^{\infty} \mbox{e}^{-V(x)/D}\;dx.
\end{equation}
Let us first consider the case of absence of the external flux, $\lambda = 0$%
. If in the noiseless case the system possesses only one stationary solution
$x_s = 0$, the probability density (\ref{s0}) has maximum at $x=0$ and the
mean value of the flux $\langle x\rangle=0$. If in the noiseless case the
system possesses three stationary states, the probability density (\ref{s0})
has three extremal points: two symmetric maxima which correspond to the
spontaneous fluxes and one minimum at $x=0$ which corresponds to the
unstable stationary state (see Fig. 1). Because the potential is
reflection-symmetric, $V(x)=V(-x)$, the mean value of any odd function of
the flux is zero. In particular, the mean value of the flux $\langle
x\rangle =0$ and the mean value of the current is zero as well. From this
point of view, properties of stationary states are trivial and non-zero
fluxes and currents are impossible. However, in some situations the
statistical moments are not good characteristics of the system because much
information is lost when an integration is performed calculating the
statistical moments \cite{horst}. The relevant quantity is a stationary
probability distribution which contains much more information about the
system. Is any reasonable method to determine the critical value of
temperature $T_c$ in this case? One possibility is to define the phase
transition in the following way \cite{horst,suzuki}: the phase transition
point is a value of the relevant parameter $\gamma$ of the system at which
the profile of the stationary distribution function changes drastically
(e.g. if a number of maxima of the distribution function changes) or if a
certain most probable point $x_0$ begins to change to an unstable state. In
some cases, it is indeed a good 'order parameter' of the system. For
example, from the measurements of the laser experiment
(see e.g. \cite{lett}),
one can obtain the stationary probability distribution of the laser
intensity and one can observe a phase transition according to the above
definition. In the case considered here, for sufficiently low temperatures,
thermal fluctuations are small and one expects the experimental results to
be accumulated around the {\it most probable values} of the stationary
probability distribution. It follows from (\ref{s0}) that the most probable
values of the flux corresponds exactly to the stationary states (\ref{stat})
of the system (\ref{evol}). In this sense, the properties of the system are
the same as discussed in the previous subsection. We want to emphasize that
it is correct for low temperatures because then the residence time in a
stable state is long. For higher temperature $T$, thermal fluctuations
become larger. In turn, fluctuations of the magnetic flux around the most
probable value become larger and larger and the residence time in a stable
state becomes shorter. One can guess that the spontaneous current should
vanish at temperature $T_0$ which is lower than the critical temperature $T_c
$ in the noiseless case. This is because of influence of Nyquist
fluctuations. The argumentation is the following. If the potential is
multistable then one can introduce characteristic time scales of the system.
The first characteristic time $\tau_d = 1/V^{\prime\prime}(x_s)$ describes
decay within the attractor $x_s=\pm x_m$ of the potential $V(x)$. The second
characteristic time is the escape time $\tau_e$ from the well around $\pm
x_m $. This time is related to the mean first passage time from the minimum
of the potential to the maximum. If these time scales are well separated,
i.e. if $\tau_e >> \tau_d$ then the description based on the most probable
value seems to be correct. Otherwise, this description fails and we should
characterize the system by averaged values of relevant variables. The exact
formula for the escape time $\tau_e$ is known \cite{gardiner} but is not
reproduced here. We have calculated $\tau_e$ for the transition from the
left minimum of the potential (\ref{pot}) assuming that the left boundary at
$-\infty$ is reflecting and the right boundary at the maximum $x_M$ of the
potential is absorbing.

\begin{figure}
\vspace{6cm} \includegraphics{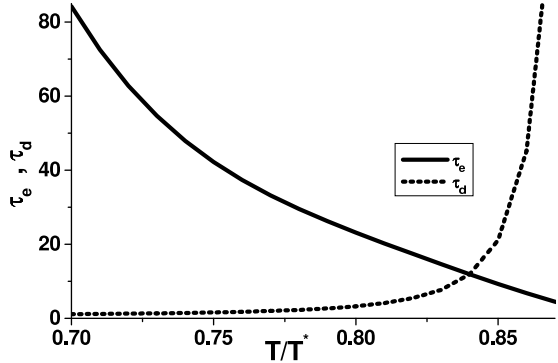} \caption{ Plot of the
characteristic escape time $\tau _e$ and decay time $\tau _d$ as a
function of temperature $T/T^*$ for $i_0=1$, $\lambda =0$ and the
strength of Nyquist noise $D=0.001T/T^*$. }
\end{figure}

In Fig. 5 we show the dependence of two
characteristic times $\tau_e$ and $\tau_d$ upon the rescaled temperature $%
T/T^*$. The escape time $\tau_e$ is a monotonically decreasing function of
temperature while the decay time $\tau_d$ monotonically increases with
temperature. In the noiseless case, for values of parameters chosen in Fig.
5 and from Eq.(\ref{pot}) we estimated the critical temperature $T_c \approx
0.88 T^*$. One can observe that roughly for temperatures $T<0.8 T^*$, the
characteristic time $\tau_d$ is more than one order of magnitude less than $%
\tau_e$. Both time scales are well separated and selfsustaining currents are
long-living states. In this sense, they are not destroyed by Nyquist noise.

Temperature $T$ enters into the Langevin equation (\ref{aa}) in two-fold: it
contributes to the intensity $D=k_BT/\epsilon _0$ of  Nyquist noise and to
the effective interaction between the rings via the potential (\ref{pot})
describing coherent electrons current. The influence of temperature on the
properties of equilibrium distribution is plotted in Fig. 6.

\begin{figure}
\vspace{6cm} \includegraphics{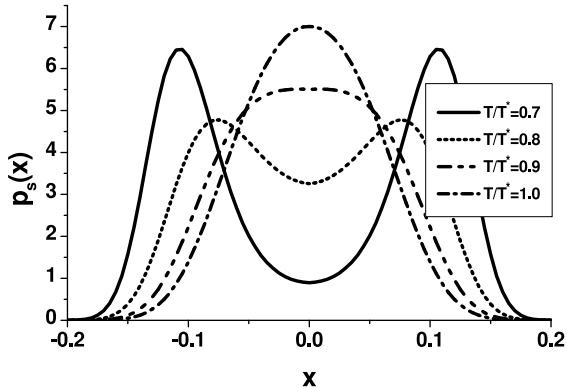} \caption{ Probability density
(28) for selected values of temperature, fixed $i_0=1$, $\lambda
=0$ and  $D=0.001T/T^*$. }
\end{figure}

 As stated
before the position of the most probably values correspond to stable
stationary states. The statistically averaged current vanishes since $%
\langle x\rangle =0$. The stationary flux variance or mean-squared deviation
$\sigma = \langle x^2 \rangle - \langle x\rangle^2 = \langle x^2 \rangle$ is
a non-monotonic function of temperature (Fig. 7): For $T=0$ the variance $%
\sigma = x_s^2$, where $x_s$ is a stationary solution of (\ref{evol}). As
the temperature increases, $\sigma$ diminishes attaining a minimal value at
some temperature $T_1$. The temperature $T_1$ seems to be always larger than
$T_c$ what is confirmed in the numerical studies.

\begin{figure}
\vspace{6cm} \includegraphics{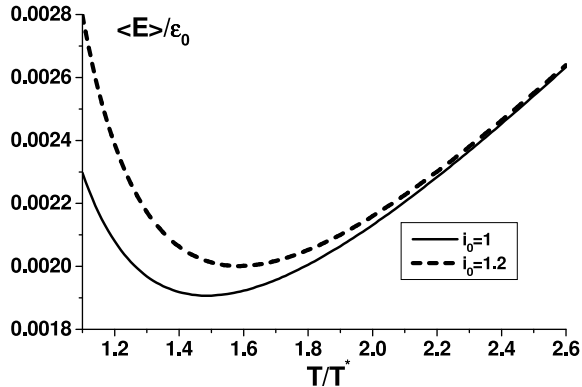} \caption{ Averaged magnetic
energy $\langle E\rangle/\epsilon _0$ given by Eq.(31) vs scaled
temperature for two values of $i_0$, fixed $\lambda =0$ and
$D=0.001T/T^*$. }
\end{figure}

 A further increase of
temperature leads to an increase of the variance. In the high temperature
limit, the dependence is linear as for the Gaussian distribution. Indeed,
below the critical temperature, the distribution (\ref{s0}) possessing two
peaks is clearly non-Gaussian. However, for higher temperatures the
probability density is one-peaked. For this case, the kurtosis
\begin{equation}  \label{kur}
Kurt = \frac{\langle x^4 \rangle}{3\langle x^2 \rangle^2}-1
\end{equation}
measures the relative flatness of the distribution (\ref{s0}) to the
Gaussian distribution. The kurtosis is negative and it means that the
distribution (\ref{s0}) is flat. It approaches zero in the high temperature
limit and then the distribution (\ref{s0}) approaches the Gaussian
distribution.

The behavior of the second moment $\langle x^2\rangle $ has a simple
explanation in terms of the average energy stored in the magnetic field, i.e
\begin{equation}  \label{energia}
\langle E\rangle = \langle\phi^2\rangle/2L=\epsilon _0 \langle x^2\rangle,
\end{equation}
where $\epsilon _0$ is given in (\ref{DV}). For low temperatures,
fluctuations are small and the main contribution to the energy comes from
the deterministic part $\phi^2/2L$. Because the magnetic flux $\phi$
decreases as temperature increases (cf. Fig. 2), hence $\langle E\rangle$
decreases as well. On the other hand, for high temperature the stationary
probability density approaches the Gaussian distribution and in consequence
the main contribution to the average magnetic energy comes from thermal
energy, $\langle E\rangle \propto kT$ which obviously increases when $T$
grows. The competition between these two mechanisms leads to the minimal
value of $\langle E\rangle$ for a certain value of temperature $T_1$.

The influence of the external field on the properties of the stationary
density (\ref{s0}) may be deduced from Fig. 3. Finally, let us consider the
limit of a very weak coupling between the ring currents corresponding to a
very small value of $i_0$. The selfsustaining stable solutions are non
accessible.  The solutions of Eq.(\ref{evol}) correspond then to the
persistent currents driven by the external field. The stationary density
forms a family of one peak curves with the most probable values given by $%
\lambda$. We conclude that even in the weak coupling limit the presence of
Nyquist noise does not destroy the persistent currents.

\subsection{Model with quenched disorder}

There are several sources of disorder which can be described as quenched
fluctuations.  The first one is caused by impurities distributed randomly in
the cylinder. The presence of impurities leads to modification of the
coherent current amplitude \cite{Riedel}. Another source of disorder is of a
geometrical origin.  The radius of rings being not exactly of the same value
may change  randomly from one ring to the other. Also the number $N_r$ of
current channels in the direction of the cylinder radius can be, in general,
different in every point of vertical axis of the cylinder. These sources of
disorder can be included into our model assuming that deviations of the
coherent current amplitude from the average are small and
\begin{equation}  \label{qu}
i_0 = j_0 + \epsilon \xi,
\end{equation}
where $j_0$ is an averaged value of the amplitude $i_0$, $\epsilon$
characterizes intensity of quenched fluctuations, $\xi$ is a zero-mean
random variable of values in the interval $[-1, 1]$ and of the probability
density $P_{\xi}(z)$.

\begin{figure}
\vspace{6cm} \includegraphics{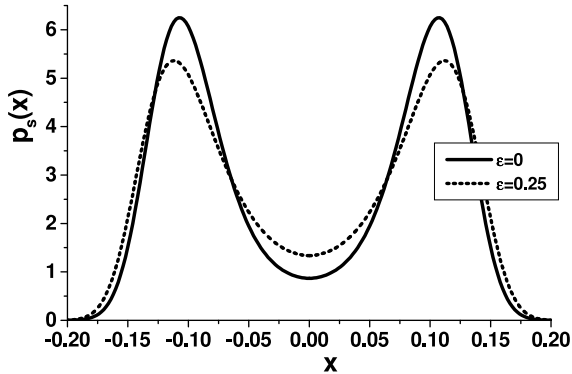} \caption{ Probability density
(33) for the system with ($\varepsilon =0.25$) and without
($\varepsilon =0$) quenched disorder for $T/T^* =0.7$, $j_0=1$,
$\lambda =0$ and $D=0.001T/T^*$ .}
\end{figure}

The stationary probability density $p_{s}(x)$ of the flux is now expressed
as
\begin{equation}  \label{sta}
p_{s}(x) = \int_{-1}^1 p(x\vert z) P_{\xi}(z) \; dz,
\end{equation}
where the conditional probability distribution
\begin{equation}  \label{cond}
p(x\vert z) =C_0(z)\; \mbox{e}^{ -[V(x) -\epsilon z F(x)]/D}
\end{equation}
has the same form as (\ref{s0}) with the replacement $i_0= j_0 + \epsilon z$
in the potential (\ref{pot}). Now, the normalization constant depends on $z$%
, namely,
\begin{equation}  \label{C0}
C_0^{-1}(z) = \int_{-\infty}^{\infty} \mbox{e}^{ -[V(x) -\epsilon z F(x)]/D}
\; dx.
\end{equation}
In Fig. 8 we show the distribution (\ref{sta}) for $p=1/2$,  $\lambda =0$ and
the uniform probability distribution $P_{\xi}(z)$. The observed shift of the
most probable value of the flux is rather small and can be interpreted as
an increase of the magnitude of the selfsustaining current. The magnitude of
peaks does not change significantly as well. One observes that both the
depth of the minimum and the height of the maxima decrease. One concludes
that the quenched disorder does not drastically change the properties of the
flux with only one exception. The probability density in or near the
critical temperature $T_c$ may change qualitatively from one peak to two
peaks. In the critical region the quenched disorder lowers the critical
temperature $T_c$.


\section{Summary}

Persistent and selfsustaining currents are beautiful manifestation
of quantum coherence in mesoscopic systems. The natural question
is how do they behave in the presence of dissipation and
fluctuations. Assuming the two fluid model for mesoscopic system
we have investigated the influence of Nyquist noise and quenched
disorder on these currents. Our discussion is limited to
stationary states of the magnetic flux and current although the
proposed model of the flux dynamics can be, in principle, applied
to study time dependent problems.

The deterministic system, obtained as a formal noiseless limit, exhibits
critical behavior such as bifurcations (phase transitions) and hysteresis.
There is a critical temperature $T_{c}$ below which non-zero selfsustaining
currents flow. For the noisy system, we have investigated the  properties of
the flux probability. We have compared the statistical properties of the
stationary probability density with the deterministic case. Maxima of the
stationary probability density correspond exactly to the deterministic
stable solutions while the local minima to the unstable solutions. In the
presence of randomness, there is a problem of defining the critical
temperature. We have presented three points of view which are based on the
mean value of the flux (no phase transitions), on the maxima of the
probability density (the same critical properties as in the deterministic
case) and on the comparison of two characteristic times (the critical
temperature lower than in the deterministic case).

The interesting finding is that the mean magnetic energy as a function of
the temperature is not monotonic - for low temperatures it is decreasing
while for high temperature it is an increasing function. The minimum is not
at the critical temperature $T_{c}$ but at temperature $T_{1}>T_{c}$.
Moreover, we have included quenched disorder and concluded that the system
is stable with respect to such a perturbation, although the most probable
values of the stationary probability density increase. It might be
interpreted as an enhancement of the current under the influence of the
quenched disorder. Therefore it serves as one more example of the
constructive role of fluctuations. The possibility of the noise-induced
current in mesoscopic rings has been considered in \cite{krawcow,mohanty}.
The presented approach can be applied not only to analysis of currents in
mesoscopic cylinders but also in superconducting rings. There is also
another class of mesoscopic systems where our model can be applied - carbon
nanotubes. The possibility of persistent currents in carbon nanotubes has
been investigated in \cite{magda}.

The general conclusion is that persistent and selfsustaining currents
survive in the presence of the above mentioned fluctuations. However, if the
intensity of Nyquist noise or quenched disorder is sufficiently strong they
lead to the lowering of the temperature below which the system is in the
ordered state. For smaller noise intensity the influence of fluctuations on
coherent currents is very small and it is very favorable for the
experimental observations of persistent and selfsustaining currents.

\section{Appendix}

For the paper to be self-contained, we remind one of the form of the
fluctuation-dissipation theorem and the Nyquist relation exploited in our
basic Eq. (\ref{lang1}). The Brownian motion of a particle of mass $m$ in a
fluid of temperature $T$ is described by a Langevin equation \cite{risken}.
According to the fluctuation-dissipation theorem \cite{risken}, its form for
the velocity $v=v(t)$ reads
\begin{equation}  \label{brow}
m \dot v + \gamma v = \sqrt{2\gamma k_B T} \; \Gamma(t),
\end{equation}
where a dot denotes a derivative with respect to time, $\gamma$ is the
friction coefficient, $k_B$ is the Boltzmann constant and $\Gamma(t)$ is the
zero-mean and Dirac $\delta$-correlated Gaussian stochastic process (white
noise),
\begin{eqnarray}  \label{gauss}
\langle\Gamma(t) \rangle = 0, \quad \langle\Gamma(t)\Gamma(s)\rangle =
\delta (t-s).
\end{eqnarray}
Mutatis mutandis, the Langevin equation for the current $I=I(t)$ in the $RL$
circuit takes the form \cite{gardiner}
\begin{equation}  \label{R-L}
L \dot I + R I = \sqrt{2 R k_B T} \; \Gamma(t).
\end{equation}
It is one of the form of the Nyquist relation. In the case when
\begin{equation}  \label{phi}
\phi = L I
\end{equation}
it can be rewritten as
\begin{equation}  \label{R-L-fi}
{\frac{1}{R}} \frac{d \phi}{dt} + \frac{1}{L}\phi = \sqrt{\frac{2 k_B T}{R}}
\Gamma(t)
\end{equation}
which justifies the prefactor of the noise term in Eq. (\ref{lang1}).

\section*{Acknowledgment}

The work supported by the KBN Grant 5PO3B0320.


\end{document}